\documentclass[traditabstract]{aa}
\usepackage{graphicx} 
\usepackage{txfonts} 

\begin{document}

\title{Three--dimensional simulations of turbulent convective mixing in ONe and CO classical nova explosions}

\author{Jordi Casanova \inst{1}
   \and Jordi Jos\'e \inst{2,3}
   \and Enrique Garc\'\i a--Berro \inst{4,3}
   \and Steven N. Shore \inst{5} }

\offprints{J. Jos\'e}

 \institute{Physics Division, Oak Ridge National Laboratory, P.O. Box 2008, Oak Ridge, TN 37831--6354, USA\
            \and
            Departament de F\'\i sica, EUETIB,
            Universitat Polit\`ecnica de Catalunya, 
            c/Comte d'Urgell 187, 
            E--08036 Barcelona, 
            Spain\
            \and 
            Institut d'Estudis Espacials de Catalunya, 
            c/Gran Capit\`a 2--4, 
            Ed. Nexus--201, 
            E--08034 Barcelona, 
            Spain\
            \and
            Departament de F\'\i sica,
            Universitat Polit\`ecnica de Catalunya, 
            c/Esteve Terrades 5, 
            E--08860 Castelldefels, 
            Spain\
	    \and
            Dipartimento di Fisica ``Enrico Fermi'',
            Universit\`a di Pisa and INFN, Sezione di Pisa, 
            Largo B. Pontecorvo 3, I--56127 Pisa, Italy
      \email{jordi.jose@upc.edu}}
       
\date{\today}

\abstract{Classical novae  are thermonuclear explosions that  take place in the envelopes of accreting white dwarfs in 
           binary  systems.   The
          material  piles   up  under  degenerate   conditions,   
          driving a thermonuclear  runaway. The energy released by the
          suite of  nuclear processes operating at  the envelope heats
          the material up to peak temperatures of $\sim (1 - 4) \times
          10^8$ K.  During these events,  about $10^{-3} -  10 ^{-7}$
          M$_{\sun}$,  enriched  in  CNO  and, sometimes,  other  intermediate--mass
          elements (e.g., Ne, Na, Mg, Al)  are  ejected  into  the interstellar  medium.  To
          account for the  gross observational properties of classical
          novae  (in  particular,  the large concentrations of metals
          spectroscopically inferred in  the ejecta),  models require mixing
          between  the  (solar--like)  material  transferred  from  the
          secondary and the outermost  layers (CO-- or ONe--rich) of the
          underlying white dwarf.
         Recent multidimensional simulations have demonstrated that 
          Kelvin--Helmholtz instabilities can naturally produce self--enrichment of
the accreted envelope with material from the underlying 
white dwarf at levels that agree with observations. However,
the  feasibility of this mechanism 
           has  been explored in the framework of CO
          white dwarfs,  while mixing with different substrates still needs 
          to be properly addressed.
         Three--dimensional simulations of 
          mixing at the core--envelope interface during nova outbursts 
          have been performed with the multidimensional code FLASH, 
          for two types of substrates: CO-- and ONe--rich. 
         We  show that the presence of an ONe--rich substrate, as in ``neon novae'',
          yields larger metallicity enhancements in the ejecta, compared to CO--rich substrates
          (i.e., non--neon novae). A number of requirements and constraints
for such 3--D simulations (e.g., minimum resolution, size of the computational
domain) are also outlined.}


\titlerunning{3--D simulations of mixing in novae} 
\authorrunning{J. Casanova et al.} 

\maketitle

\section{Introduction}

Classical novae repeatedly eject $\sim 10^{-3} - 10^{-7}$ $M_\odot$ 
into the interstellar medium, on intervals of years 
(for the fastest recurrent novae) to tens of millennia (for the most common 
classical novae). 
High--resolution spectroscopy  has revealed the presence of 
nuclear--processed  material relative to solar abundances in these ejecta
 (see, e.g., Gehrz et al. 1998, Downen et al. 2013).  Nova explosions are thought to be 
 important sources of Galactic $^{15}$N,
$^{17}$O, and $^{13}$C (see Starrfield, Iliadis and Hix 2008, 2016,  Jos\'e and Shore 2008, 
and Jos\'e 2016, for recent reviews).   Hydrodynamic simulations have shown
that nuclear processing of the accreted material alone is unable to account for 
the large metallicities inferred in the ejecta because of the moderate peak temperatures reached 
during the outburst. Instead, 
 mixing at the core--envelope interface has been regarded as a feasible
alternative to nuclear processing. Several mechanisms have been proposed
to this end, including diffusion--induced convection (Prialnik and Kovetz 1984;
 Kovetz and Prialnik 1985; Iben et al. 1991, 1992; Fujimoto and Iben 1992), 
shear mixing (Durisen 1977; Kippenhahn and Thomas 1978; MacDonald 1983; 
Livio and Truran 1987; Kutter and Sparks 1987; Sparks and Kutter 1987), 
convective overshoot--induced flame propagation (Woosley 1986), and 
convection--induced shear mixing (Kutter and Sparks 1989). While all
 these mechanisms likely participate in producing the  metallicity 
enhancement of the ejecta, 1--D simulations have shown that none can 
successfully account for the full range of values inferred spectroscopically
(see Livio and Truran 1990, for a detailed account of the shortcomings
of these different mechanisms).

It has become 
increasingly clear that 1--D, spherically symmetric simulations,
while capable of reproducing the main observational features of the nova outburst 
exclude an entire sequence of events associated with the way a thermonuclear 
runaway initiates and propagates (Shara 1982).   With the advent of supercomputing capabilities in the past decades, 
a number of 2--D and 3--D 
simulations of mixing at the core--envelope interface during nova outbursts
have been published  (Glasner and Livne 1995, 
Glasner, Livne, and Truran 1997, 2005, 2007; Kercek, Hillebrandt, and Truran 1998, 1999;
Rosner et al. 2001; Alexakis et al. 2004; Casanova et al. 2010, 2011a,b).
These  multidimensional efforts have confirmed that Kelvin--Helmholtz instabilities 
that appear at the late stages of the explosion can naturally lead to self--enrichment 
of the accreted envelopes with material from the underlying (CO) white dwarfs, 
at levels consistent with observations.   The results are robust, being independent of the specific choice of the initial 
perturbation (duration, strength, location, and size), numerical resolution,  or the size of the computational domain (Casanova et al. 2011a).

To date, only two studies have reported multidimensional (2--D) simulations with
different compositions for the underlying white dwarf
(Glasner, Livne, and Truran 2012, 2014). As in previous work, only a slice of the star (0.1
$\pi$ rad), in spherical--polar coordinates (with a maximum resolution
 of 1.4 km $\times$ 1.4 km), was modeled,  with
 a reduced network containing 15 isotopes
(up to $^{17}$F; Glasner et al. 2012) and 35 isotopes (up to $^{27}$Al;
Glasner et al. 2014). The simulations assumed a 1.147 $M_\odot$
 white dwarf and different compositions (CO, pure He, pure $^{16}$O, and pure 
$^{24}$Mg).  The studies support the idea that Kelvin--Helmholtz instabilities also operate 
for different chemical substrates. However, no realistic multidimensional 
simulation of mixing for the typical composition of an ONe white dwarf were
reported in Glasner et al. (2012, 2014). The present paper aims to fill this
gap.
 We report the first 3--D simulation of mixing at the core--envelope 
interface during nova outbursts for ONe white dwarfs.

The paper  is organized  as follows. The input physics and initial  
conditions of the simulations reported are described in Sect. 2. 
A full account of two 3--D simulations of mixing performed at the core--envelope
interface of an ONe white dwarf and, for  comparison, a CO--rich white dwarf,
are presented in Sect. 3. Finally, the significance of the results presented and
the main conclusions are summarized in Sect. 4.

%-------------------------------------------------------

\section{Input physics and initial setup}
Accretion of solar composition material ($Z \sim 0.02$) onto a 
1.25 $M_\odot$ ONe white dwarf, at a rate of $2 \times 10^{-10}$ $M_\odot$ yr$^{-1}$, was 
 simulated with the one--dimensional, implicit, Lagrangian hydrodynamic code SHIVA 
(Jos\'e \& Hernanz 1998, Jos\'e 2016).  We have used this code extensively for the modeling of classical 
nova outbursts, from accretion to the ultimate explosion, expansion, and ejection stages. 
We adopted the composition of the outer white dwarf layers from Ritossa, Garc\'\i a--Berro, and Iben (1996): 
X($^{16}$O) = 0.511, X($^{20}$Ne) = 0.313, X($^{23}$Na) = 0.0644, X($^{24}$Mg) = 0.0548, 
X($^{25}$Mg) = 0.0158, X($^{27}$Al) = 0.0108, X($^{12}$C) = 0.00916, X($^{26}$Mg) = 0.00989, 
X($^{21}$Ne) = 0.00598, and X($^{22}$Ne) = 0.00431.  When the temperature at the core--envelope interface reached $T_{\rm ce} \sim 10^8$ K, the structure was mapped onto a three--dimensional cartesian grid
 (hereafter, Model A) which was subsequently followed with the multidimensional, parallelized, explicit, Eulerian FLASH code.    It is based on the piecewise parabolic interpolation of physical quantities to solve the hydrodynamic equations that describe the stellar plasma (Fryxell et al. 2000).   The code uses adaptive mesh refinement to improve accuracy in critical regions of the computational 
domain.  The 3--D computational domain\footnote{Radially, the envelope extends 325 km above the ONe core, in Model A. 
The outer 75 km of the underlying white dwarf have also been included in the computational domain.}, 
800 $\times$ 800 $\times$ 400 km$^{3}$, initially comprised 88 unevenly spaced radial layers and 
512 equally spaced layers along each horizontal axes, in hydrostatic equilibrium. The maximum resolution adopted, with 5 levels of refinement,  was  1.56 $\times$ 1.56 $\times$ 1.56 km$^{3}$ to handle the 
sharp discontinuity at the core--envelope interface, although the typical zoning employed in 
the simulation was 3.125 km in each dimension. Periodic boundary 
conditions were implemented at lateral sides, while hydrostatic conditions 
were imposed through 
the vertical boundaries, reinforced with a reflecting condition at the bottom and an outflow 
condition at the top (Zingale et al. 2002; Casanova et al. 2010, 2011a,b). 
A reduced nuclear reaction network containing 31 species 
($^{1}$H, $^{4}$He, $^{12,13}$C, $^{13,14,15}$N, $^{14,15,16,17}$O, $^{17,18}$F, $^{20,21}$Ne, 
$^{21,22,23}$Na, $^{22,23,24,25}$Mg, $^{24,25,26g,26m,27}$Al, $^{26}$Mg, and $^{26,27,28}$Si, 
where $^{26g}$Al and $^{26m}$Al represent the ground and a short--lived isomeric state in $^{26}$Al,
respectively) was included to treat the energetics of the event.  The network was linked through 41 nuclear interactions, mainly p--captures and $\beta^{+}$--disintegrations.  The corresponding rates were taken from the STARLIB nuclear 
reaction library (Iliadis et al. 2010).    At the first timestep, to break the initial equilibrium configuration, we introduced a top--hat temperature perturbation (5\%)   in a 4 km wide spot, located at $(x, y, z) = (5 \times 10^{7}, 5 \times 10^{7}, 3.754 \times 10^{8})$ cm, close to the core--envelope interface. 

For comparison, we computed a second three--dimensional model of mixing with a CO--rich substrate (hereafter, Model B).   As in Model A, the early accretion and explosion stages were  computed in spherical symmetry with the SHIVA code,
assuming accretion of solar composition material onto a 1 $M_\odot$ CO 
white dwarf at a rate of $\sim$ $2 \times 10^{-10}$ $M_\odot$ yr$^{-1}$.  
 The total envelope mass accreted in Model B reached 
$6.2 \times 10^{-5}$ $M_\odot$ ($2.2 \times 10^{-5}$ $M_\odot$ in Model A).
 The reduced nuclear 
network was the same as in Casanova et al. (2010, 2011a,b) since the main nuclear activity does not 
extend beyond the CNO--region. This network contained 13 chemical species 
($^{1}$H, $^{4}$He, $^{12,13}$C, $^{13,14,15}$N, $^{14,15,16,17}$O, and 
$^{17,18}$F) linked through 18 nuclear interactions.  The composition of the underlying white dwarf was 
X($^{12}$C) = X($^{16}$O) = 0.5. The initial configuration  and the temperature perturbation were chosen to match the same initial conditions as Model A.   

Both 3--D simulations used 512 Intel SandyBridge processors at the MareNostrum supercomputer (BSC), 
requiring about 90 khr of computation for Model A  and 25 khr for Model B.  

\section{Results}

In both models,  the initial temperature perturbation creates fluctuations along the core--envelope interface that develop strong 
buoyant fingering. The early development of these primary fluid instabilities  
is depicted in Fig. 1; we show the $^{20}$Ne (Model A; upper panel) and $^{12}$C mass fractions (Model B; lower panel). Movies
 showing the development of Kelvin--Helmholtz instabilities, 
 up to the time when the convective front hits
the upper computational boundary, are available online or at {\tt http://www.fen.upc.edu/users/jjose/Downloads.html}.

Fingering appears within the first 50 seconds. However, for the ONe--rich substrate (Model A) 
the development of such fluid instabilities continues for a longer time, up to $\sim$ 170 s.  This  
drives an efficient dredge--up of outer core material into the envelope by the rapid formation of small convective 
eddies at the innermost envelope layers. In Model A, the build--up of the first circulation (convective) 
structures occurs at $\sim$ 200 s, much later than for Model B (i.e., $\sim$ 50 s). 
Ignition for solar--metallicity models\footnote{See,
however, Shen and Bildsten (2009) for ignition conditions in C--poor envelopes.}
is driven by the reaction $^{12}$C+p, which is faster than the alternative channels, $^{16}$O+p or $^{20}$Ne+p.
Therefore, the amount of $^{12}$C present in the envelope critically determines the ignition time.
Since the presence of a C(O)--rich substrate (as in Model B) produces larger dredge--up of $^{12}$C into the
envelope than for an ONe--rich substrate, ignition (and in turn, the establishment of superadiabatic gradients 
required for convection) occurs earlier in Model B.  

\begin{figure}[h!]
\includegraphics[scale=0.23]{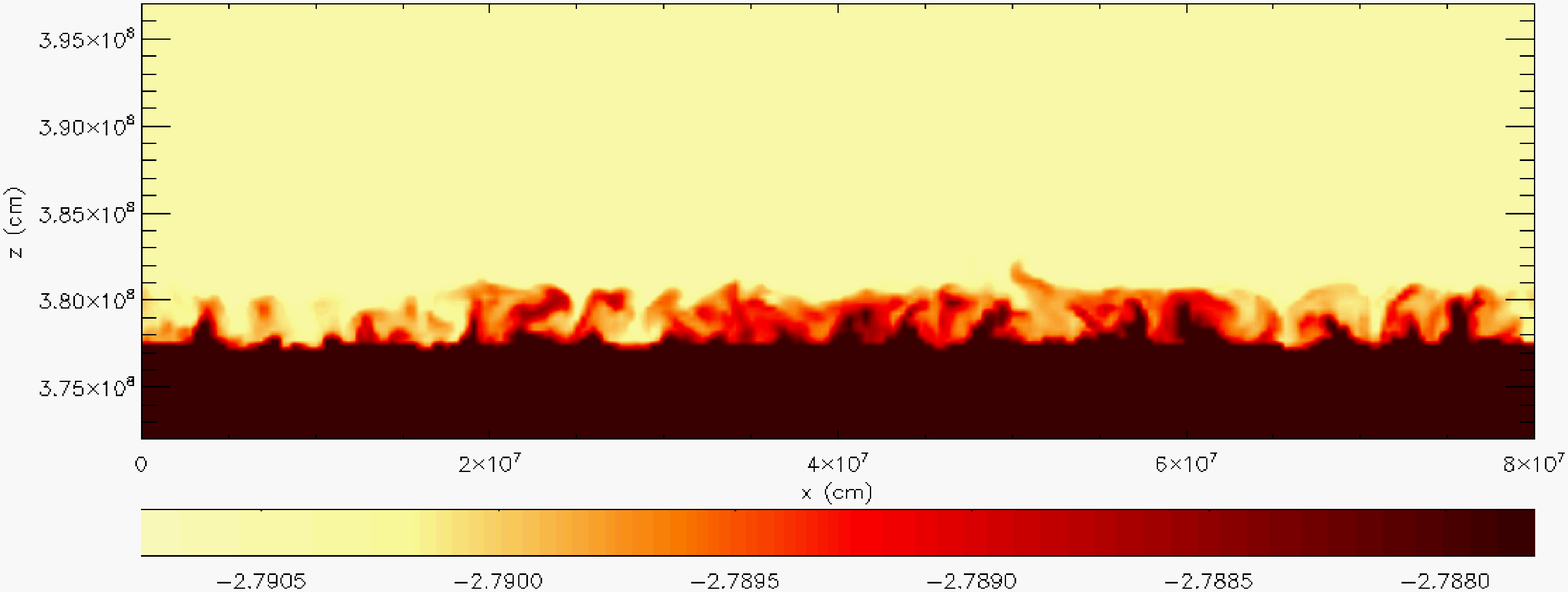}
\includegraphics[scale=0.23]{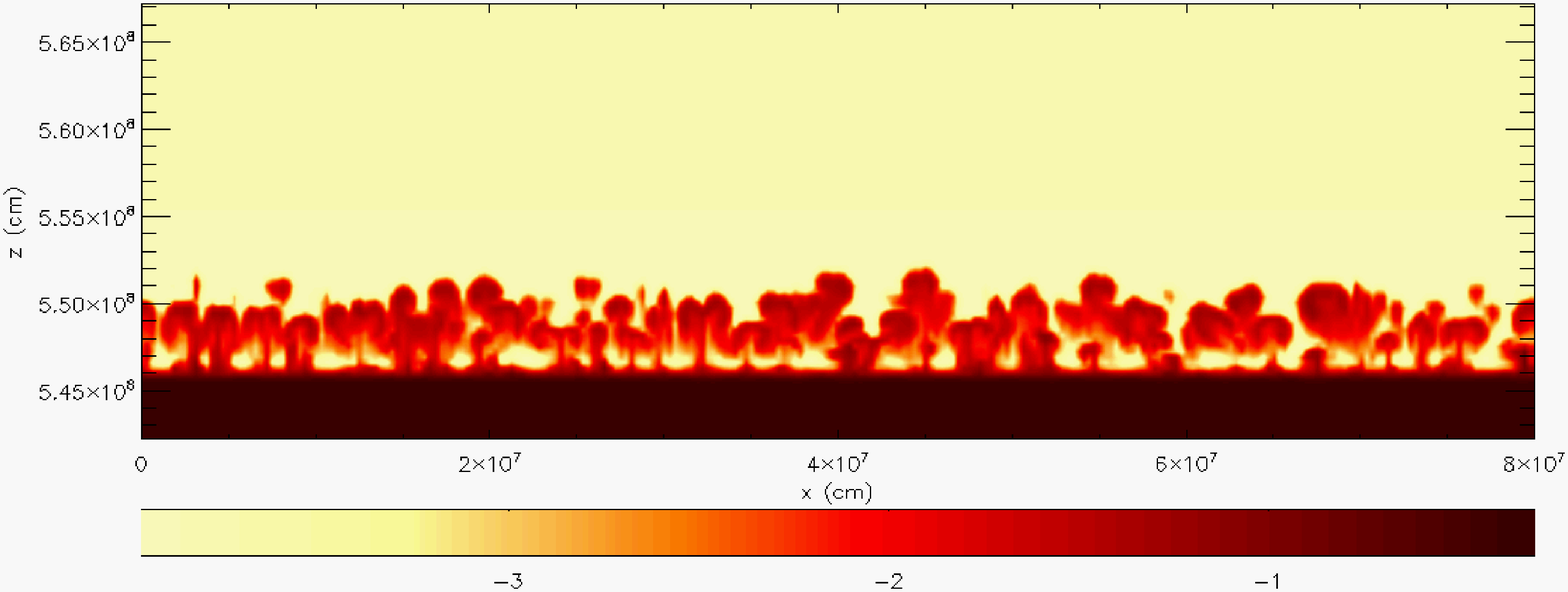}
\caption{Snapshots of the early development and growth of buoyant fingering for Model A ($t = 130$ s; upper panel) 
and Model B ($t = 40$ s; lower panel), shown in terms of  
$^{20}$Ne and $^{12}$C mass fractions, respectively, in logarithmic scale.
Both snapshots correspond to 2--D slices on the $xz$ plane, taken at $y = 4 \times$ $10^{7}$ cm. In both models, the 
initial buoyant fingering appears within the first 50 s, although it lasts longer (up to 170 s) in Model A.}
\end{figure}

The onset of convection is accompanied by shear flows at the core--envelope interface. 
This causes dredge--up of outer core material by means of Kelvin--Helmholtz instabilities. As a result, 
the envelope becomes progressively enhanced 
in intermediate--mass nuclei. As time goes on, the convective front extends throughout the entire envelope (Fig. 2).
In fact, the expansion and progress of the explosion towards the outer envelope layers proceeds 
almost in spherical symmetry. This explains the success of one--dimensional models in describing the main
observational features of classical nova outbursts (Starrfield et al. 1998, 2009; Kovetz $\&$ Prialnik 1997; 
Yaron et al. 2005; Jos\'e $\&$ Hernanz 1998). 
As burning proceeds, the velocity and the temperature of the material increase. After about 500 s, convection turns strongly intermittent and, apparently, turbulent.  This state continues to the end of the computation.  The flow exhibits structures on a wide range of scales,  with filaments as large as 200 km that extend from the core--envelope interface through the upper envelope layers.    The flow shows indications of a cascade, with the largest eddies spawning progressively smaller features that, in turn, fragment and merge.  The interface becomes more irregular after $\sim$ 600 s.  
Individual knots and filaments of enhanced $^{20}$Ne abundance persist on scales of about 50 km for up to $\sim$ 10 s.  
In contrast to the static display from about 490 s to about 560 s, 
a rapid advance of the mixing 
front\footnote{There may be sufficient time during this vigorous 
convective stage to build a fast dynamo.
If so, the magnetic field could amplify before the onset of the explosion 
and be advected with the layer. Such magnetic fields can become very large 
since buoyancy is inhibited by the rapidly expanding layers. This may have 
important consequences for the evolution of those systems (e.g., 
ultra-high energy production [$>$ 100 MeV gamma-rays] during particle 
acceleration).}
begins at about 560 s.  The Ne front jumps from 392 km to 404 km in about 37 s.   At this stage, burning proceeds 
as a subsonic flame.  This is the same conclusion reached in Casanova et al. (2011b) for the CO case.   This filamentary behavior and the development of vortex structures of different sizes are actually a signature of 
intermittency and an observational evidence of the energy transfer cascade into smaller scales 
predicted by Kolmogorov's theory of turbulence (Lesieur et al. 2001; Shore 2007; Pope 2000). 
 The turbulent flow shows density contrasts that become chemically inhomogeneous.  The nuclear yield is, therefore, characterized as a distribution rather than a sharp value, as in one dimensional models.   This is illustrated by the abundance distribution of a trace species, $^{15}$O,  in a  100 km wide layer (comparable to the pressure scale height), located well above the core--envelope interface, extending from $3.91 \times 10^{8}$ to $4.01 \times 10^{8}$ cm, 
at the beginning ($t = 0$ s) and at the end of the simulation ($t = 751$ s).
As shown in Fig. 3, the initially discrete distribution evolves into a  stable form 
that is approximately a Gaussian with a dispersion of 25\%.   There is, however,  a so--called ``fat tail'' that extends towards high abundances that accounts for  $\sim$7\%\ of the yield.  Such chemical inhomogeneities have been observed in the ejecta of many classical novae that did not suffer much interaction with the ambient material 
(Shore et al. 1997; Porter et al. 1998; Vaytet et al. 2007). Although several collisions and supersonic motions following the outburst 
can amplify the initial inhomogeneities, their physical origin points to the intermittent behavior of turbulence.

\begin{figure}[h!]
\includegraphics[scale=0.28]{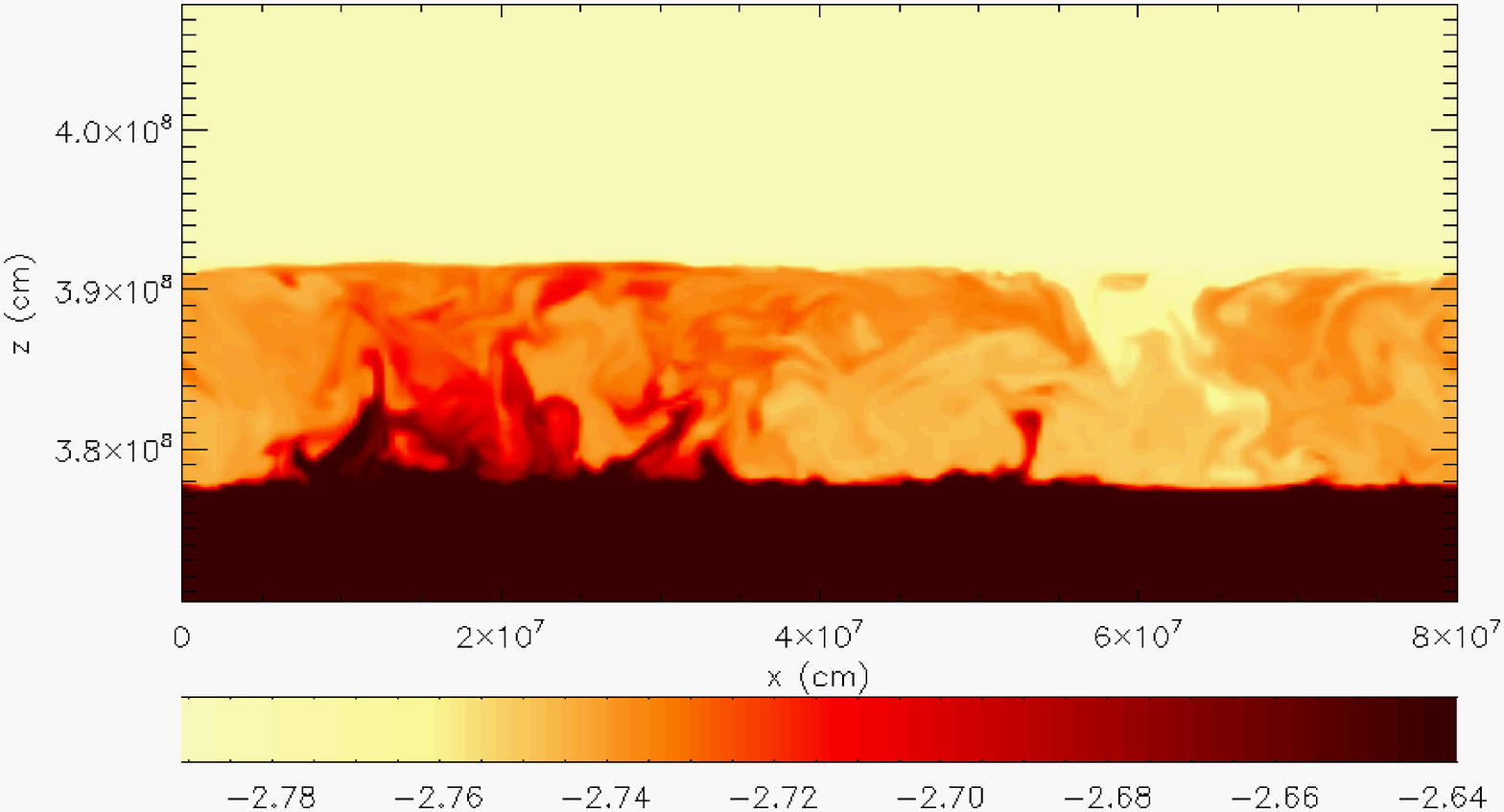}
\includegraphics[scale=0.28]{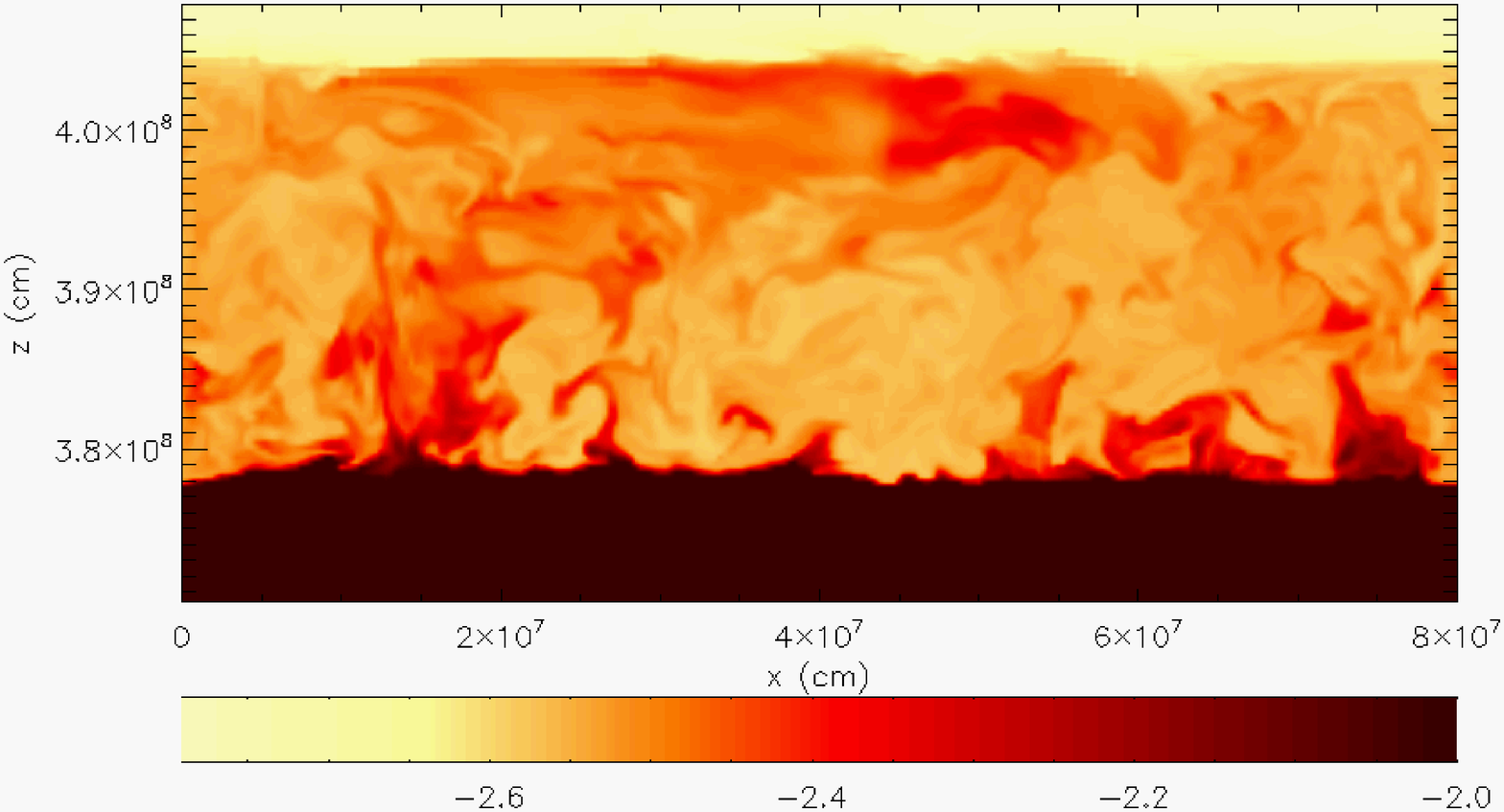}
\includegraphics[scale=0.28]{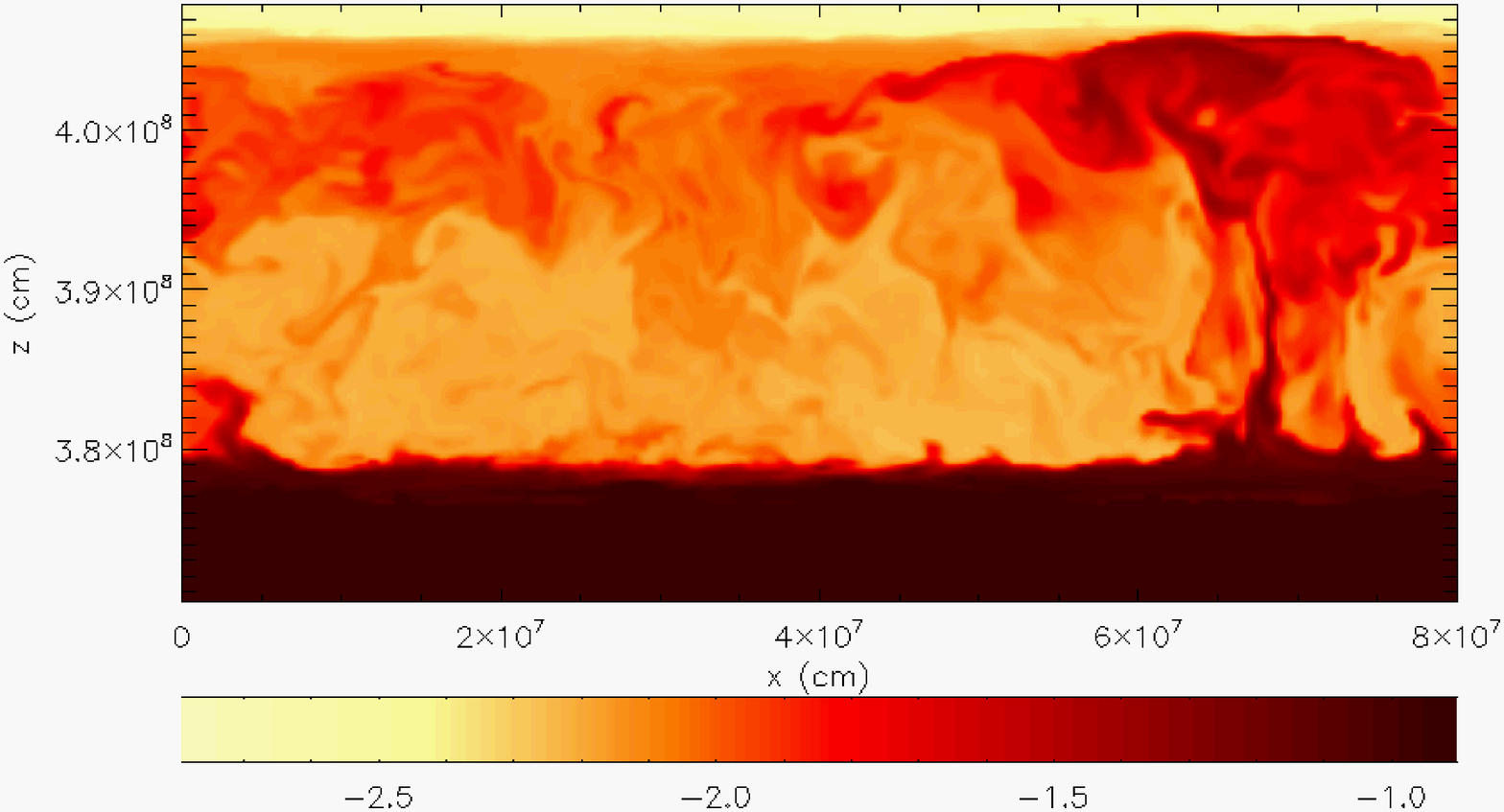}
\caption{Snapshots of the development of the convective front in Model A, at 
$t = 437$ s (upper panel), 635 s 
(central panel), and 709 s (lower panel), shown in terms of the $^{20}$Ne mass fraction, in logarithmic scale. 
Snapshots correspond to 2--D slices on the $xz$ plane, taken at $y = 4 \times 10^{7}$ cm. The dredge--up of fresh 
 material from the outer core, driven by Kelvin--Helmholtz instabilities, translates into a mean metallicity in
the envelope of 0.211, 0.234 and 0.241, respectively. The mean metallicity at the end of the simulation reaches 0.246, 
by mass, for this model.}
\end{figure}

\begin{figure}[h!]
\includegraphics[scale=0.35, angle=-90]{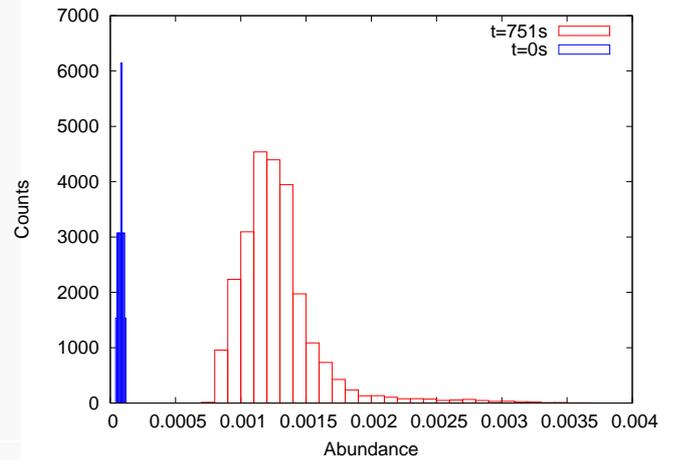}
\caption{Abundance distribution of $^{15}$O at the beginning (blue) and at the end ($t = 751$ s;
red) of the 
simulation, in Model A. The histograms contain about 24,500 points from a 100--km thick layer, 
located well above the core--envelope interface ($\sim$150 km). As time goes on, the initially narrow and 
discrete distribution evolves into an inhomogeneous distribution that fits a Gaussian with a dispersion of 25\% in 
the mean value.}
\end{figure}

Model A was stopped as soon as the convective front reached the top of the computational domain, after sweeping 325 km (corresponding to $\sim$ 3 pressure scale heights).  The maximum temperature in this simulation was $2.02 \times {10}^{8}$ K with peak fluid velocities $\sim$ ${10}^{7}$ cm s$^{-1}$, and therefore, very low Mach numbers.  At the end of the simulation, the mean metallicity of the envelope reached 0.246, about 12 times the initial value. These results confirm that Kelvin--Helmholtz instabilities in the presence of an ONe--rich substrate can also yield large metallicity enhancements in the envelope 
through convective dredge--up, at levels agreeing with observations.  For Model B, the development of the outburst and the mixing episodes proceeds similarly, but the metallicity enhancement is lower, $Z = 0.105$ (about 5 times the initial value). This results from the shorter duration of the dredge--up phase, since mixing with a $^{12}$C--rich substrate
significantly reduces the time required to drive a thermonuclear runaway.

\section{Discussion and limitations}

 The feasibility of Kelvin--Helmholtz instabilities  as a mechanism for 
self--enrichment of accreted envelopes with material from different substrates, has also been 
explored by Glasner, Livne, and Truran (2012).  They computed the 1--D evolution of an accreting 1.147 $M_\odot$ 
white dwarf, subsequently mapping the $3.4 \times 10^{-5}$ $M_\odot$ 
accreted envelope as well as the outer $4.7 \times 10^{-5}$ $M_\odot$ of 
the underlying white dwarf into a 2--D, spherical--polar grid.  
All the different substrates considered consisted of one--species gas  
(i.e., pure He, O, Ne, or Mg).  The authors found that ``significant enrichment (around 30 per cent) of the ejected layer,
by the convective dredge--up mechanism, is a common feature of the
entire set of models, regardless of the composition of the accreting
white dwarf.''  Other properties of the outburst, such as the 
characteristic timescale of the explosion, were found to depend 
more sensitively on the composition of the substrate.

A detailed comparison between our results and those of Glasner et al. (2012) faces a number of limitations. 
The one species assumption in Glasner et al. is an important oversimplification when modeling the composition of the substrate.  The white dwarfs that produce nova explosions are CO-- or ONe--rich but also contain a blend of  
different species. Moreover, the critical mass and extent 
of the accreted envelopes depend critically on the properties of underlying 
white dwarf.   The layer is more massive for CO-- than for ONe--rich white 
dwarfs, since the latter are more massive than the former 
(see, e.g., Starrfield et al. 1989, Jos\'e and Hernanz 1998, Yaron et al. 2005, and 
Jos\'e 2016), an effect not taken into account in Glasner et al. since they aimed at testing the importance of Kelvin--Helmholtz instabilities in driving mixing with different substrates rather than a quantitative 
analysis of the extent of such mixing.  In contrast, the study we present is a self--consistent simulation of mixing during a nova employing  realistic models for the ONe--rich substrate.

\begin{figure}[h!]
\includegraphics[scale=0.35, angle=-90]{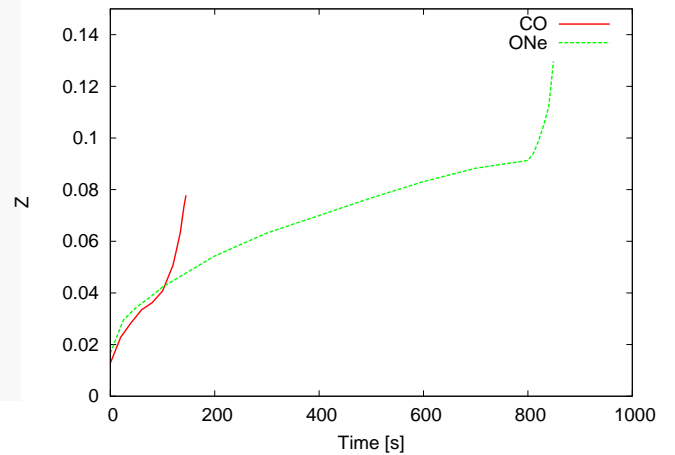}
\caption{Time--evolution of the mean metallicity of the envelope in the
2--D models with underlying CO (solid line) and ONe white dwarfs 
(dashed line).}
\end{figure}

Determining the influence of the substrate on the degree of mixing is, however, 
hard to assess directly from our models.  Using different envelope sizes and masses in our 3--D simulations, although self--consistently computed in 1--D hydrodynamics, coupled to different 
substrates makes it hard to disentangle the effects of substrate abundance, white dwarf mass, and accreted mass.  We therefore computed two 2--D high--resolution ($780 \times 780$ m) models, with the same white dwarf mass (1.15 $M_\odot$) and accreted 
envelope ($3.2 \times 10^{-5}$ $M_\odot$; 350 km), to specifically address the effect of the composition of the underlying substrate.  The increase with time of the mean envelope metallicity is shown in 
Fig. 4. A higher metallicity was reached with an ONe--rich  ($Z \sim 0.129$) 
than with a CO--rich substrate ($Z \sim 0.078$).  Although an extensive analysis of mixing with ONe vs. CO white dwarfs would
require a large set of multidimensional simulations for different masses (a project we are currently undertaking), the current 
paper demonstrates that the longer duration of the thermonuclear runaway, caused by the lower $^{12}$C content in an ONe--rich substrate, translates into a longer dredge--up phase that results  in a higher degree of mixing. 

A final issue, the difference in final mean envelope metallicity between the 2--D and 3--D models with an ONe substrate, deserves further analysis.    It may result, at least in part, from the different adopted white dwarf masses since their critical  envelope masses and sizes also differ.  Additionally, the final envelope metallicities reported here are actually lower limits, since the simulations were stopped before envelope--core detachment occurred.   Finally, some of the difference may be caused by  
numerical diffusion,  arising from the different resolution adopted in the 2--D and 3--D simulations.  To perform such a survey is, however,  very computationally expensive and quite taxing of currently available resources.

In summary, the 3--D simulations reported here indicate that mixing with  
$^{12}$C--poor substrates  (e.g., an ONe white dwarf; Model A) lead to larger metallicities in the envelope, and in turn, in the ejecta. 
 Some cautionary words must be added in this regard, however. The two multidimensional simulations reported in this work were initiated (mapped) when the temperature at the envelope base reached $10^8$ K, neglecting the possible contribution of early mixing occurring at previous stages of the thermonuclear runaway. The possible effect of such early mixing has been investigated, however, by Glasner et al. (1997, 2007) in 2--D,  for different choices of the initial temperature, demonstrating that all models converge to an almost universal model. Indeed, the convergence in the mixing amounts in those models, including a case initiated at $7 \times 10^7$ K, remained within 5\%. Such conclusions  need to be verified in 3--D. 

{\it Acknowledgements.} 
The authors would like to thank Alan Calder, for many fruitful discussions and exchanges.
Part of the software used in this work was 
developed by the DOE--supported ASC/Alliances Center for Astrophysical
Thermonuclear Flashes at the University of Chicago. This work has been 
partially
supported by the Spanish MINECO grant AYA2014--59084--P,
 by the E.U. FEDER funds, by the AGAUR/Generalitat de Catalunya grant 
SGR0038/2014, and by the U.S. Department of Energy Office of Nuclear Physics.
We also acknowledge the Barcelona Supercomputing
Center for a generous allocation of time at the MareNostrum supercomputer.

\end{document}